\documentclass[aps,prd,nofootinbib,twocolumn,floatfix]{revtex4-1}
\usepackage{graphicx}
\usepackage{dcolumn}
\usepackage{subfigure}
\usepackage{times,mathptm}
\usepackage{float}
\usepackage{color}
\usepackage{amsmath,amsfonts}
\usepackage{mathptmx}
\usepackage{mathrsfs}
\usepackage{bbm}
\usepackage{bm}
\usepackage{xfrac}

\newcommand{\beq}{\begin{eqnarray}}
\newcommand{\eeq}{\end{eqnarray}}

\newcommand{\D}{\mathcal{D}}

\renewcommand{\d}{\delta}
\renewcommand{\l}{\lambda}
\renewcommand{\L}{{\cal T}}
\renewcommand{\b}{\beta}

\newcommand{\tr}{\text{Tr}}

\newcommand{\m}{\mu}

\renewcommand{\r}{\rho}

\renewcommand{\D}{\Delta}

\newcommand{\oh}{\frac{1}{2}}

\newcommand{\dg}{\dagger}
\newcommand{\non}{\nonumber}

\newcommand{\rf}[1]{(\ref{#1})}
\newcommand{\ra}{\rightarrow}

\usepackage{ulem}

\begin{document}
\bibliographystyle{h-physrev5}

\title{Adjoint torelons, and the persistence of color \\
 electric flux tubes in the deconfined phase} 

\author{J. Greensite$^1$ and {\v S}. Olejn\'{\i}k$^2$}
\affiliation{$^1$ Physics and Astronomy Dept., San Francisco State
University, San Francisco, CA~94132, USA \\
$^2$ Institute of Physics, Slovak Academy of Sciences, SK--845 11
Bratislava, Slovakia}

\date{\today}
\begin{abstract}
 
    It is argued that the adjoint torelon loop, i.e.\ a Polyakov loop in the adjoint representation running
in a spatial, rather than temporal, direction, is an observable which is sensitive to the presence
of long color electric flux tubes at high temperatures.   We show via lattice Monte Carlo simulations that this observable has a sharp peak at the deconfinement transition, remains much larger than the vacuum value for some range 
of $T>T_c$, and falls below the vacuum value for $T > 2T_c$.   This result suggests that
long electric flux tubes may persist for a finite range of temperatures past the deconfinement transition, and at some stage disappear, presumably melting into a plasma of gluons.  As a side remark, we point out that our results at $T<T_c$ imply that the eigenvalues of ordinary Polyakov loop holonomies in the confinement phase have a slight tendency to attract rather than repel, which may be relevant to certain models of confinement. 
\end{abstract}

\pacs{11.15.Ha, 12.38.Aw}
\keywords{Confinement,lattice
  gauge theories}
\maketitle

\section{\label{sec:intro}Introduction}

    The thermal average of an operator $O$ in a canonical ensemble is given by the standard formula of
quantum statistical mechanics
\beq
\langle O \rangle_{thermal} = {\sum_n \langle n | O |n \rangle e^{-E_n/kT} \over \sum_n  e^{-E_n/kT}     } 
\label{sm}
\eeq
where $\{|n\rangle\}$ is the set of energy eigenstates.  In the case of a pure gauge theory, two questions naturally
arise:  First, at low temperatures, the thermal average is dominated by the ground state, so what is this 
ground state, exactly?  The question could be answered, e.g.\ in the Schrodinger representation, by specifying the Yang-Mills vacuum wavefunctional $\Psi_0[A] = \langle A | 0 \rangle$.   Secondly, at temperatures greater than the deconfinement temperature $T_c$,  what are the relevant excited states that dominate the thermal average?
If these relevant states $|n\rangle  = Q_n |0 \rangle$ at $T>T_c$ can be constructed by acting on the ground state with certain gauge-invariant operators $\{Q_n\}$, then what is the form of such operators?

    We are probably very close to having an answer to the first question, at least in $D=2+1$ directions.  It is likely,
for the reasons given in refs.\ \cite{Greensite:2007ij,Greensite:2011pj}, that the Yang-Mills vacuum wavefunctional
has the form
\beq
\Psi_0[A] &=& \exp\left[-{1\over 2g^2}\int d^2x d^2y ~ F_{12}^a(x) \right.
\non \\
 & & \left. \qquad \times  \left({1 \over \sqrt{-D^2 - \l_0 + m^2}} \right)^{ab}_{xy} F_{12}^b(y) \right] \; ,
\label{vacuum}
\eeq
or at least something quite close to that, where $D^2$ is the covariant Laplacian in the adjoint representation, $\l_0$ is the lowest eigenvalue of $-D^2$, and $m$ is a mass parameter.\footnote{An alternative possibility is that the kernel has instead the form
$1/(\sqrt{-D^2 -\l_0+m^2}+m)$, which reduces to the Karabali, Kim, Nair proposal \cite{Karabali:1998yq} when evaluated on abelian configurations.  The two kernels are qualitatively very similar, and the corresponding wavefunctionals are difficult to distinguish quantitatively in numerical tests, cf.\ \cite{Greensite:2011pj}.}   

   The situation is much less clear at high temperatures, just past the deconfinement transition.   All that can really be said is that in temporal gauge the relevant $Q_n$ must be gauge-invariant operators, but this fact provides only modest guidance.  Apart from such ultralocal operators as $\tr[F^2]$, gauge-invariant operators can be built out of combinations of the eigenvalues and eigenstates of gauge-covariant operators.  For example, if ${\cal D}$ is a Dirac operator, say, or a covariant Laplacian, and ${\cal D} \phi_n  = \l_n \phi_n$ is the corresponding eigenvalue equation, then any functional
of the set of functions $\{\phi^a_n(x) \phi^a_m(x)\}$, and the eigenvalues $\{\l_n\}$, is gauge-invariant.  The exponent in eq.\ \rf{vacuum} is another example.  Gauge-invariant functionals can also be built from transformations to a physical gauge.  Let $g[x;A]$ be the transformation which takes a gauge-field $A_k$ into Coulomb gauge.  Then any functional $G[g[x;A] \circ A_k(x)]$ is also a gauge-invariant operator. 

   There are two types of operators, however, which may play a special role in pure Yang-Mills theory:  the Wilson loops 
$W(C)$, which create thin lines of electric flux, and the 't Hooft loops $B(C)$, which create thin lines of magnetic flux known as center vortices.   Unlike the previous examples, these operators do not require, as part of their definition, the solution of a differential equation over the full spatial volume.  The low-lying energy eigenstates of strong-coupling lattice gauge theory are calculable, and can certainly be expressed in terms of Wilson loop operators acting on the vacuum state.  There is also strong evidence that center vortices, created by the `t Hooft loop operator, play an important role in the confinement mechanism \cite{Greensite:2003bk}, including the generation of a spacelike string tension at high temperatures \cite{Engelhardt:1999fd}. So one guess is that the relevant excited states at high temperature are created by products and superpositions of these two special types of operators, i.e.
\beq
 |n\rangle &=& \sum \{\mbox{electric flux tube operators} \} 
\non \\
&  & \times \{\mbox{magnetic flux tube operators}\}  |0\rangle \; .
\label{e&m}
\eeq
Chernodub, Nakamura, and Zakharov \cite{Chernodub:2008iv} have presented evidence that a network of magnetic flux tubes in the deconfined phase is responsible for large contributions to relevant thermodynamic quantities in the
hot Yang-Mills equation of state.

   In this article we will be mainly interested in the electric flux tube content of the plasma state.  At low temperatures, well below $T_c$, the relevant $Q_n$ will simply create low-lying glueball states, which are understood to be small closed electric flux tubes.  As $T$ increases, longer electric flux tubes (highly excited glueball states), are less suppressed, and become increasingly relevant.  It is a very old idea that flux tubes percolate at $T=T_c$; the flux tube free energy goes to zero while the energy per unit length remains finite.   We are interested in the question of what happens at $T>T_c$.   Do the color electric flux tubes simply vanish?  Or do they persist for some range of high temperatures, before ``melting" into a plasma of gluons?

\section{\label{sec:torlon} Adjoint torelons and electric flux tubes}

   In this section we will explain why the thermal average of adjoint torelon lines, i.e.\ adjoint Polyakov loops in a spatial direction, might be relevant to the study of electric flux tubes at finite temperature.    From the start we should confess that the argument we will present is entirely heuristic.   Nevertheless, this argument serves to motivate our calculation of the adjoint torelon line, and it will suggest an interpretation of the results.

\begin{figure*}[t!]
\begin{center}
\subfigure[]  
{   
 \label{glump-a}
 \includegraphics[scale=0.80]{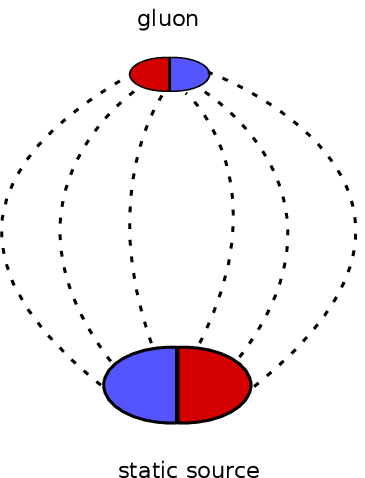}
}
\hspace{1.5cm}
\subfigure[]   
{  
 \label{glump-b}
 \includegraphics[scale=0.80]{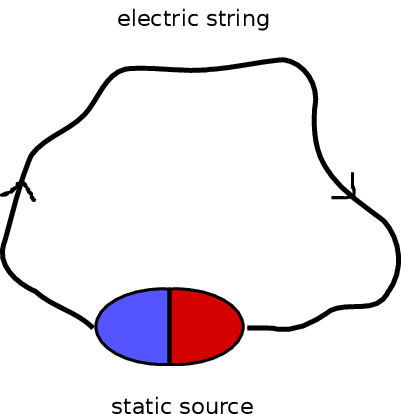}
}
\hspace{1.5cm}
\subfigure[]   
{  
 \label{glump-c}
 \includegraphics[scale=0.80]{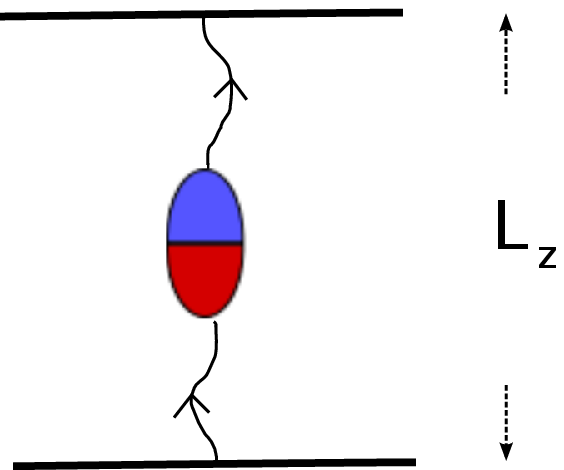}
}
\end{center}
\caption{Descriptions of the gluelump state:  (a) The gluelump as a bound state of a static source and a constituent
gluon in Coulomb gauge.  Dashed lines indicate the color Coulomb interaction.  The red/blue shadings correspond to the adjoint representation as a two-index object.  (b) The gluelump as a fluctuating electric string pinned to the static source.  (c) For spatial extension $L_z$ comparable or smaller than the gluelump size, the electric string will tend to wind around the periodic volume.}
\label{glump} 
\end{figure*}

   We begin by considering Polyakov loops in the adjoint representation, of extension $L_t$ in the time direction.  As is well known, adjoint Polyakov loops have a non-zero vacuum expectation value even in the confined phase, because the color charge of a static adjoint source can be screened by gluons.  The resulting state is known as a ``gluelump," and it can be pictured in two ways.   First, in some appropriate (e.g.\ Coulomb) gauge, a gluelump can simply be thought of as a bound state consisting of the static source and a single constituent gluon, Fig.\ \ref{glump-a}.  Secondly, just as a glueball can be pictured as a closed, fluctuating string of electric flux, a gluelump can be thought of as a closed electric string which begins and ends at the static source, Fig.\ \ref{glump-b}.  
   
   The evidence that a closed electric flux tube can be described by a fluctuating string has become quite compelling in recent years, for both 2+1 and 3+1 dimensions.  In particular, lattice Monte Carlo simulations have found that the low-lying spectrum of closed flux tube states, which are closed by lattice periodicity, is described to a surprising degree of accuracy by the simple Nambu-Goto action.  The ground state energy in particular agrees with the Casimir energy of the Nambu string (associated with the L\"uscher term) down to tube lengths where one might think that a simple string picture of the flux tube should not apply.  Up-to-date results can be found in refs.\ 
 \cite{Athenodorou:2011rx,Athenodorou:2010cs}.
We will therefore be guided by the picture of a glueball as a closed, fluctuating string, and assume that this picture also applies to gluelumps, which would then resemble strings attached to static D0-branes.  A gluelump has some finite extension which, like that of most low-lying hadrons, is probably on the order of one fermi.   This rough estimate is supported, to some extent, by the fact that the string between two static adjoint sources breaks into two gluelumps when the sources reach a separation of 1.25 fm, as determined from lattice Monte Carlo studies \cite{deForcrand:1999kr}.   The length of the ``broken" string attached to one of the sources, at the breaking point, serves as a crude upper bound for the radius of the gluelump.

    Now consider what will happen as the extension $L_z$ of the periodic volume in the spatial $z$-direction is reduced to or below the extension of the gluelump.  In the constituent gluon picture, the bound gluon can take advantage of the periodicity in the $z$-direction by having zero momentum in this direction; this reduces the kinetic energy of the bound gluon without increasing color charge separation beyond the normal extension of the gluelump.  The ground state gluelump wavefunction is then completely delocalized in the $z$-direction.  Likewise, in the string picture, the string can wind around the periodic $z$-direction, as indicated in Fig.\ \ref{glump-c}, without incurring a large cost in energy.  In the string picture, as $L_z$ becomes very small, the string partition function is dominated by states with multiple windings, and the contribution from states with zero winding number is negligible by comparison.   In the case of the gluelump, where the ends of the string are pinned to a source and the center of mass cannot propagate freely in the $z$-direction, the winding mode represents an extra degree of freedom, strongly suppressed at large $L_z$, which opens up at small $L_z$.
    
    In string theory, of course, the $L_z \ra 0$ limit of a D0-brane is equivalent, by T-duality, to the $L_z\ra \infty$ limit for a D1-brane.  In the latter case, summation over the center-of-mass momentum in the $z$-direction contributes a factor of $L_z$ to the thermal partition function (see, e.g., Zwiebach \cite{Zwiebach:2004tj}), so the D1-brane partition function diverges linearly with $L_z$.  Likewise, the sum over winding modes of the D0-brane leads to a divergence as $L_z \ra 0$.   Of course, the expectation value of an adjoint Polyakov line at finite temperature is not necessarily the same thing as the thermal partition function of string ending on a D0-brane.  Nevertheless, since the low-temperature flux tube spectrum is so well described by the Nambu action, we are led to expect that string winding modes will enhance the expectation value of the adjoint Polyakov loop, as $L_z$ decreases.
  
    Next let us interchange the $z$ and $t$ labels, so that the ``short" direction is now the time direction, and the Polyakov loop runs along the large spatial $z$-direction.   Polyakov loops which run in a spatial direction are sometimes referred to as ``torelons," and we will adopt that terminology here.    The expectation value of an adjoint Polyakov loop in a 4-volume with one short space direction can be reinterpreted as the thermal expectation value of an adjoint torelon loop at finite temperature.   The string pictures of the gluelump state, Figs.\ \ref{glump-b} and \ref{glump-c}, now take on a somewhat different meaning.

\begin{figure}[thb]
\begin{center}
\subfigure[]  
{   
 \label{section-a}
 \includegraphics[scale=0.45]{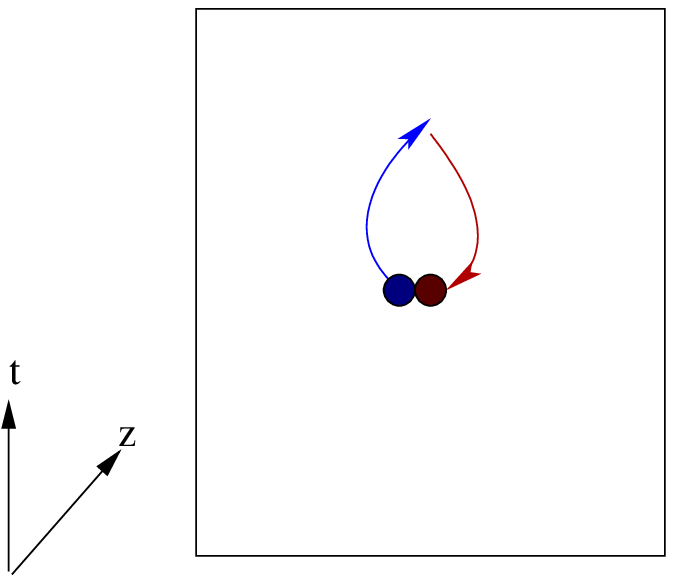}
}
\hspace{1.0cm}
\subfigure[]   
{  
 \label{section-b}
 \includegraphics[scale=0.45]{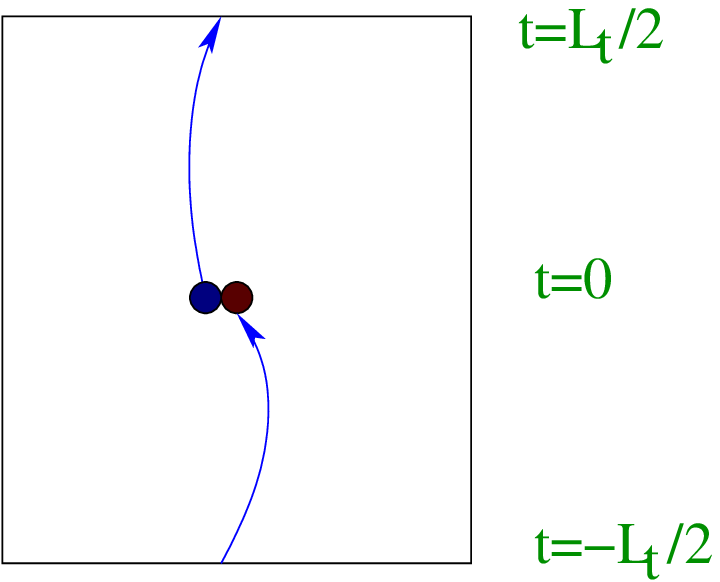}
}
\end{center}
\caption{Schematic diagram of string propagation in a space-slice at fixed $z$.  (a) Creation at $t=0$ of an electric string-antistring pair, which subsequently annihilate.  (b)  An electric string created at $t=0$ which propagates through the periodic lattice, and is destroyed at $t=0$.}
\label{section} 
\end{figure}

\begin{figure}[tbh]
\begin{center}
\subfigure[]  
{   
 \label{strong-a}
 \includegraphics[scale=0.40]{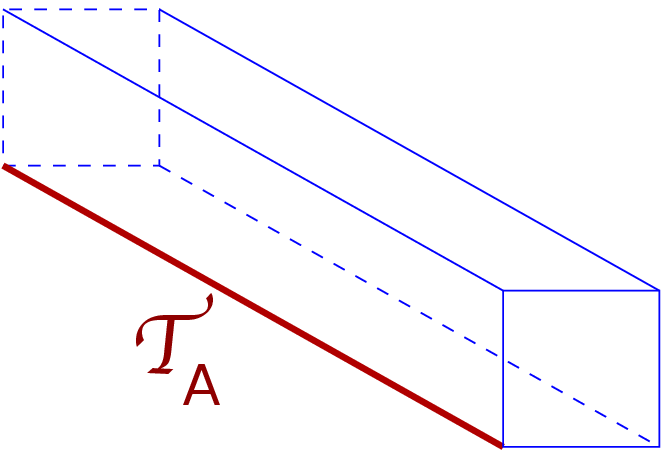}
}
\hspace{1.5cm}
\subfigure[]   
{  
 \label{strong-b}
 \includegraphics[scale=0.40]{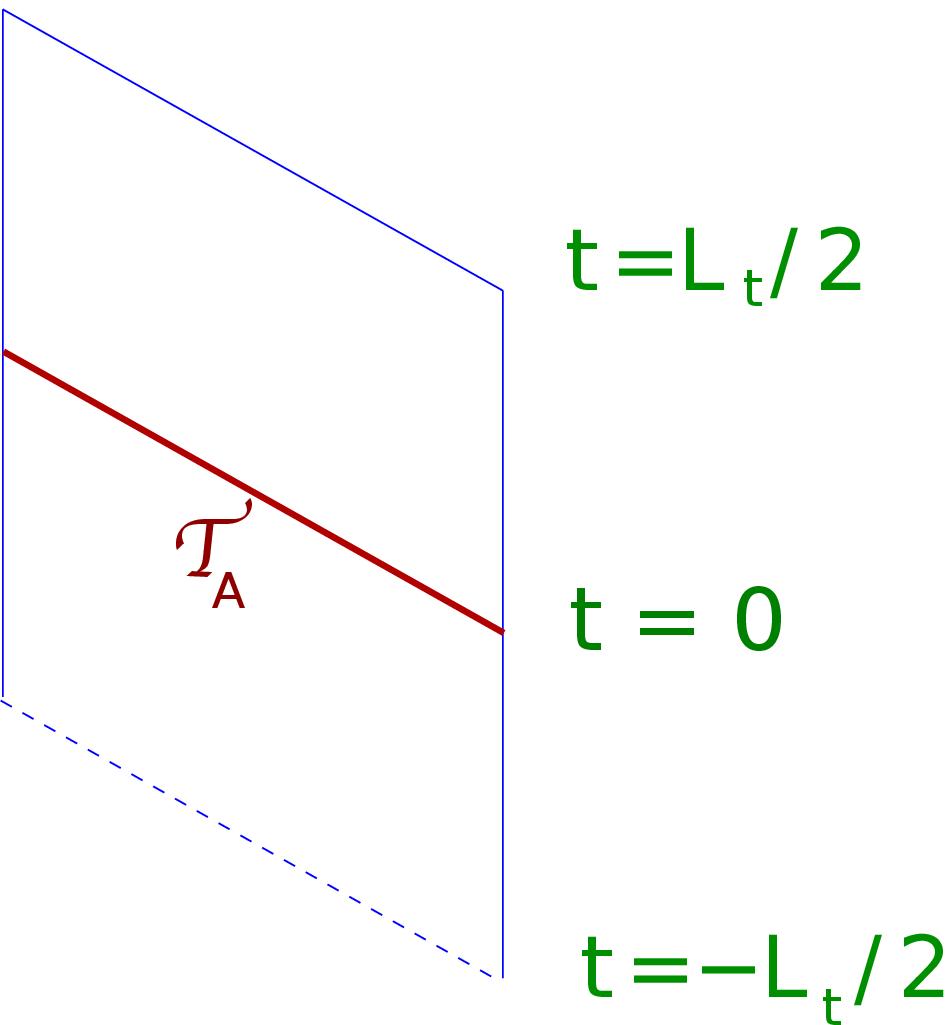}
}
\end{center}
\caption{Lattice strong-coupling diagrams corresponding to (a) Fig.\ \ref{section-a}; and (b) Fig.\ \ref{section-b}.}
\label{strong} 
\end{figure}


\begin{figure*}[t!]
\centerline{\scalebox{0.8}{\includegraphics{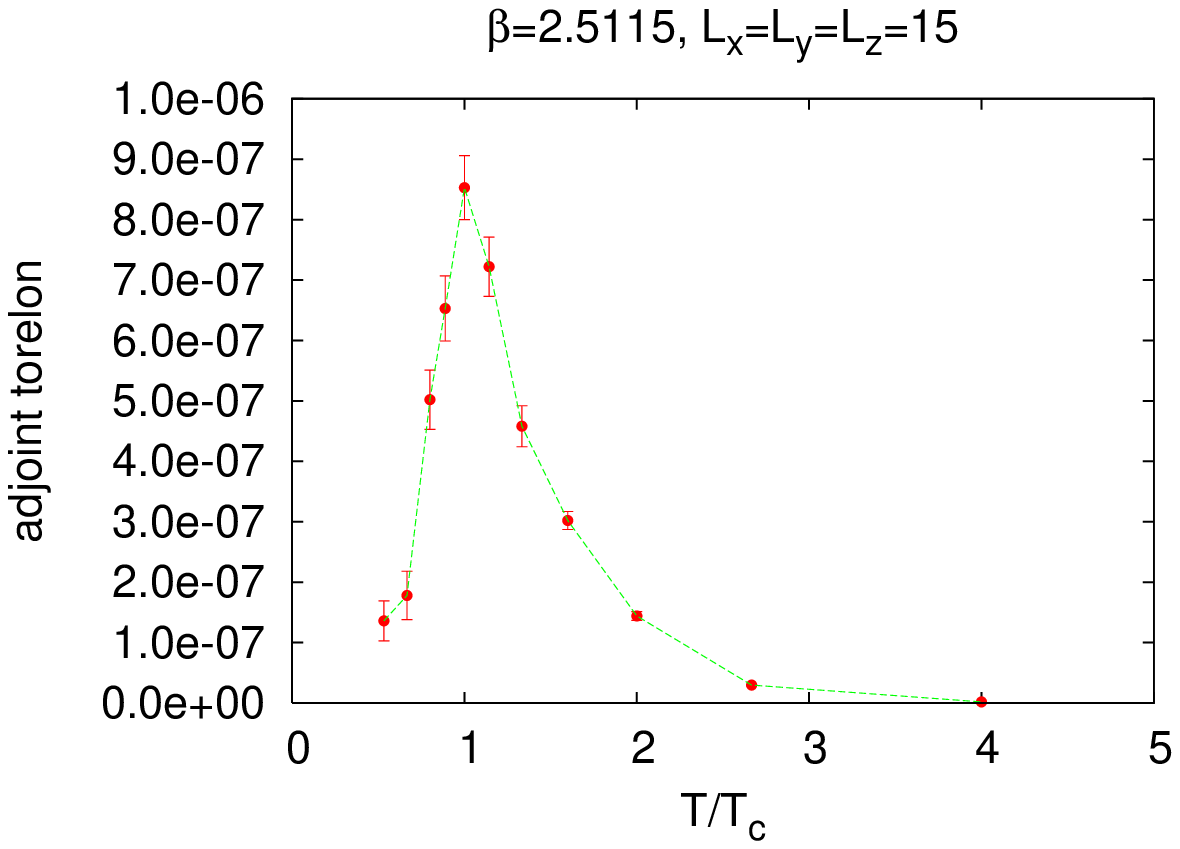}}}
\caption{Thermal average of the adjoint torelon line $\langle \L_A \rangle_{thermal}$ on a $15^3 \times L_t$ lattice, vs.\ temperature ratio $T/T_c$.  The deconfinement temperature at $\b=2.5115$ corresponds to $L_t=8$ lattice spacings in the time direction, and therefore $T/T_c = 8/L_t$. The dashed line is inserted to guide the eye.}
\label{plas251}
\end{figure*}
    
     Before going into that, it is useful to first note that physical states in an SU($N$) pure gauge theory can be characterized by their transformation properties with respect to an underlying global center symmetry defined at any fixed time.  In a lattice formulation, in a $t=0$ timeslice, consider the global transformation
\beq
           U_z(x,y,z=0) \ra  e^{i 2\pi / N}  U_z(x,y,z=0)   \; ,             
\label{center}
\eeq
with all other links at $z\ne 0$ unchanged.  A physical state $\Psi[U]$ has $N$-ality $k$ if, under the global transformation \rf{center},
\beq
            \Psi[U] \ra  e^{i 2\pi k/ N} \Psi[U] \; .
\eeq
An operator which changes the $N$-ality of a state by $\D k = 1$ is the torelon operator in the fundamental representation
\beq
\L = {\rm Tr~P}\exp\left[ig\int_0^{L_z} dz~  A_z(00z0)\right] \; ,
\eeq
or in lattice regularization
\beq
\L = {\rm Tr}\left[\prod_{n_z=1}^{N_z} U_z(00{n_z}0)\right] \; ,
\eeq
while the conjugate operator $\L^\dg$ changes the $N$-ality by $\D k = -1$.
Any superposition of Wilson lines
\beq
          \sum_C a(C)  {\rm Tr~P}\exp\left[ig\oint_C dx^\m A_\m \right] 
\eeq
which are closed by lattice periodicity in the $z$-direction, with winding number
equal to one, also change the $N$-ality of states by one unit.   A thick flux tube which winds once through
the periodic lattice is created by operators of this type, acting on the vacuum. 
The vacuum state itself, assuming ordinary periodic boundary conditions, has $N$-ality zero, and in fact the
lattice version of the vacuum wavefunctional \rf{vacuum} is explicitly invariant under the transformation
\rf{center}.  

    Because $\L$ and $\L^\dg$ take states from one $N$-ality sector to another, we can regard them as creating/destroying long electric strings which wind through the periodic lattice in the $z$-direction.  Operator $\L$ creates a string or destroys an ``antistring,"  i.e.\ a string with the opposite orientation of color electric flux, while $\L^\dg$ creates an antistring or
destroys a string. The adjoint torelon loop is simply a Polyakov loop which runs in the $z$-direction, in the adjoint representation of the gauge group, i.e.
\beq
\L_A =  \L \L^\dg  - 1 \; .
\eeq

Now the string diagrams for gluelumps, Figs.\ \ref{glump-b} and \ref{glump-c}, under the $z \leftrightarrow t$ interchange, can be reinterpreted as follows:  At low temperatures, with $L_t$ much greater than the gluelump extension, 
Fig.\ \ref{glump-b} can be viewed as the creation of a string-antistring pair, which propagate for a finite time and subsequently annihilate.  This process is indicated schematically in Fig.\ \ref{section-a}, which shows a slice of the volume at fixed $z$; i.e.\ the adjoint torelon line runs into the page.  The corresponding diagram in lattice strong-coupling perturbation theory is shown in Fig.\ \ref{strong-a};  it involves a sheet of plaquettes wrapping around the adjoint line to form a tube.  It is contributions of this type that contribute to the vacuum expectation value of the adjoint torelon loop.  At high temperatures, with $L_t$ less than the typical gluelump extension of roughly one fermi,  the winding string in Fig.\ \ref{glump-c} becomes relevant.  This figure can be reinterpreted, as in Fig.\ \ref{section-b}, as a string worldsheet which wraps around the periodic time direction, passing through the $z$-axis at $t=0$.  The corresponding lattice strong-coupling diagram is shown in Fig.\ \ref{strong-b}; in this case the sheet of plaquettes wraps around the periodic time direction.  As explained in, e.g., ref.\ \cite{Chernodub:2006gu}, a particle worldline which wraps around the periodic volume in the time direction can be interpreted as a real particle in Minkowski space, rather than a virtual particle in Euclidean space, which contributes to the thermal ensemble.  Likewise, the wrapped worldsheet represents contributions to $\langle \L_A \rangle_{thermal}$ due to states containing at least one long electric string, closed by periodicity in the 
$z$-direction.  
 
    The identification of hyperplanes at $t=\pm L_t/2$ corresponds, in the canonical ensemble, to the sum over states
in \rf{sm}.   We have already argued that the degree of freedom associated with the string winding mode, which opens up at small compactified dimension, ought to enhance the expectation value of an adjoint loop.  But since a winding number $k>0$ corresponds, in the thermal field theory, to states of $N$-ality $k$ mod $N$, such processes can only contribute if states  containing a minimum of $k$ long flux tubes are present in the thermal ensemble at high temperatures.  Thus, the greater the electric flux tube content in the thermal ensemble, the greater the contribution from the wrapped worldsheets, and the larger the value of $\langle \L_A \rangle_{thermal}$.   Conversely, if long electric flux tube states are for some reason almost absent in the thermal ensemble, then only worldsheets  with zero winding number and associated $N$-ality $k=0$ can contribute, as in the $T=0$  case, and all other potential contributions are suppressed.

    The qualitative argument just presented is our reason for thinking that any significant enhancement of an adjoint torelon line, beyond its $T=0$ value, may indicate the presence of states containing long electric flux tubes in the thermal ensemble.   With this motivation, we have computed the expectation value of the adjoint torelon line in SU(2) lattice gauge theory, using the standard Wilson action at  $\b=2.5115$ on lattices of volume $15^3 \times L_t$.  The extension
$L_t$ in the timelike direction runs from 15 down to 2 lattice spacings, and the torelon line, running in the $z$-direction, is always 15 lattice spacings long.   The critical temperature at $\b=2.5115$ corresponds to 8 lattice spacings in the time direction \cite{Fingberg:1992ju}, so we are sampling temperatures from a low of ${8\over 15} T_c$ to a high at $4 T_c$.  The lattice spacing at $\b=2.5115$ is $a=0.08$ fm, so if the extension of the gluelump is about one fermi, we would expect that winding contributions would become significant at around 12 lattice spacings ($T={2\over 3} T_c$).  In this calculation we have made use of the L\"uscher-Weisz noise reduction technique \cite{Luscher:2001up} with one level of sublattices, each of three lattice spacings extension in the $z$-direction.

Our results are displayed in Fig.\ \ref{plas251}.   Although the magnitudes of the adjoint torelon expectation values are always very small compared to $O(1)$, there is a dramatic structure seen in the relative values, which show a sharp peak precisely at the deconfinement transition.  According to our previous reasoning, this peak could be attributed to the presence of long electric strings  for temperatures at or near the deconfinement transition.  Two points are worth noting.  First, the adjoint loop does not immediately drop to (or below) the vacuum value at $T>T_c$.  For temperatures a little beyond $T_c$, the thermal average is still much greater than the vacuum expectation value.  On the other hand, as
$T$ increases beyond $2T_c$, $\langle \L_A \rangle_{thermal}$ falls to values which are very much smaller than the low-temperature value.  Our interpretation of these results is that states with long electric flux tubes become increasingly important as the temperature increases, up to the deconfinement transition temperature.  Beyond that transition, long electric flux tubes persist for some range of temperatures, until, at high enough temperatures, the
relevant excited states no longer contain such configurations, and the flux tubes presumably ``melt" into the surrounding plasma.

\section{Do Polyakov loop eigenvalues attract, or repel?}

Our calculation of the adjoint torelon line at the lowest temperature $T={8\over 15}T_c$, obtained on a hypercubic $15^4$ lattice volume, is also (again interchanging the $z$ and $t$-directions) a reasonable estimate of the expectation value of an adjoint Polyakov loop at this (comparatively) low temperature.\footnote{A previous calculation by Gupta et al.\ \cite{Gupta:2007ax} of the adjoint Polyakov line in SU(3) pure gauge theory was carried out in the confinement phase for temperatures quite close to the transition temperature, $T/T_c > 0.907$.  A positive value was obtained,  but for our purposes it is desirable to check this result at a much lower temperature, well below the transition point, where high temperature effects are hopefully negligible.}  This data point allows us to address quite a different question, namely:  In a pure Yang-Mills theory, in the confinement phase, do the eigenvalues of a {\it fundamental} representation Polyakov loop tend to attract, or tend to repel?   

    Obviously the endpoint of attraction is where the eigenvalues are identical, and this must be a center element, while the endpoint of repulsion is a situation where the trace of the Polyakov loop is zero.  An example of the former is $Z_N$ lattice gauge theory, where Polyakov loops are precisely center elements and the VEV vanishes, in the confined phase, due to quantum fluctuations among these center elements.  This is also the case for
center-projected SU($N$) lattices in maximal center gauge.   The situation is similar to the disordered phase of a $Z_N$ spin system where there are islands of the same center element, but upon averaging over the full volume, the elements cancel.  In the ordered phase, the lattice is overwhelmingly one center element or another. The opposite extreme is the dilute dyon gas advocated by Diakonov and Petrov \cite{Diakonov:2007nv}.  In that picture, in the confined phase, the trace of almost any Polyakov loop (away from the middle of the dyon) is zero in each configuration, which is the limiting case of eigenvalue repulsion.   Deconfinement occurs when there is a switch (at the minimum of the free energy) from eigenvalue repulsion to eigenvalue attraction.

    Of course, these pictures are extremes.  In reality the distribution of Polyakov eigenvalues is close to random in the confined phase, i.e.\ close to the Haar distribution.\footnote{In fact, even if center symmetry were explicitly broken by matter fields in the fundamental representation,  the eigenvalue distribution would still be nearly random at low temperatures, just due to non-confining fluctuations.}  The same can be said for the eigenvalue distribution of large Wilson loop holonomies.  But there is, nevertheless, a lot of information contained in the eigenvalue distribution of Wilson loop holonomies.  The tiny deviation from the Haar measure in that case encodes information about the string tension, the L\"uscher term, Casimir scaling, and color screening.  For this reason, we may expect the small deviation from the Haar measure for Polyakov loops to be significant as well. If the underlying confining configurations tend to have Polyakov eigenvalues coincide, as in a center vortex mechanism, or repel, as in a dilute dyon gas, then these tendencies might be expected to show up in the small deviation of the eigenvalue distribution away from the Haar measure. 
    
   Let 
\beq
U(L)  =  \mathrm{P}\exp\left[i\int_0^{L_t} dt A_0({\bf 0},t)\right] 
\eeq
be a Polyakov loop holonomy, at spatial coordinates ${\bf x}$ taken to be ${\bf x}={\bf 0}$, and let
$P=\mbox{Tr}[U(L)]$.  If the eigenvalues of $U(L)$ attract, then the probability distribution for $U(L)$ on the group manifold should be slightly peaked around center elements; i.e.\ $P=z N$, where $z\in Z_N$ for the SU($N$) gauge group.  Conversely, if the eigenvalues tend to repel, then the distribution will be peaked around configurations for which $P=0$.

    The probability density $\r(g)$ for a Polyakov holonomy is given by
\beq
             \r(g) = \langle \d[g-U(L)] \rangle \; .
\label{density}
\eeq
where the Haar measure distribution corresponds to $\r(g)=1$.
Now make a character expansion of the delta function, and at this point specialize to the SU(2) gauge group, so that
\beq
\d[g-U(L)] = \sum_{j=0,\oh,1,...} \chi_j[U(L)] \chi_j[g] \; .
\eeq
Then, in the confined phase
\beq
\langle \d[g-U(L)] \rangle = \sum_{j=0,1,2,3...} \langle \chi_j[U(L)] \rangle \chi_j[g] \; .
\eeq
Note that the half-integer representations are dropped in the confined phase, on grounds that
$\langle \chi_j[U(L)] \rangle$  vanishes, for $j=$ half-integer, if center symmetry is unbroken.  But, apart from the trivial $j=0$ representation, Polyakov loop VEVs are very small even for integer $j$, and are very rapidly suppressed as $j$ increases.  It is therefore an excellent approximation to the sum to keep only the leading contributions, i.e.
\beq
\r(g) = \langle \d[g-U(L)] \rangle \approx  1 + \langle \text{Tr}_A[U(L)] \rangle \text{Tr}_A[g] \; ,
\eeq
where $\text{Tr}_A[g]=\chi_1[g]$ is just the trace in the adjoint representation.

   The crucial question is now whether $\langle \text{Tr}_A[U(L)] \rangle$ is positive, or negative.  If it is positive, then the Polyakov loop probability density is largest when $g$ is a center element ($= \pm \mathbbm{1}$ for SU(2)), and we can conclude that the eigenvalues tend to attract.  Conversely, if the trace of the adjoint Polyakov loop is negative, then the
probability density is largest at $\text{Tr}_A[g] = -1$, which is obtained when the trace in the fundamental representation is zero.  In that case, the eigenvalues tend to repel.

    We have seen, from the first data point (obtained on a $15^4$ lattice) shown in Fig.\ \ref{plas251}, that the trace of an adjoint Polyakov loop at low temperatures is positive.  We conclude from this data that the eigenvalues of a Polyakov loop tend to very slightly {\it attract}, rather than repel, in the confinement phase.       

\section{Conclusions}

    If the qualitative argument in section \ref{sec:torlon} is correct, then the peak we have seen in the adjoint torelon loop at the deconfinement transition is an indication that long electric flux tubes survive, in the thermal ensemble, somewhat beyond the deconfinement transition, up to $T \approx 2T_c$.  For a true quark-gluon plasma created via heavy ion collisions, this would suggest that  some hadronic bound states, such as closed flux tubes (i.e.\ glueball states) and certain quark bound states, may survive a little beyond $T_c$.  This conclusion ties in very well with the discussion of  Shuryak in ref.\ \cite{Shuryak:2009cy}, who argues that certain aspects of recent RHIC data could be explained by the existence of metastable color electric flux tubes in the quark gluon plasma, at temperatures somewhat beyond $T_c$.  Shuryak and co-workers have also argued for the existence of flux tubes past the transition based on a dual superconductor model of confinement, cf.\ 
\cite{Faroughy:2010cd,Liao:2007mj}.   As we will show in a subsequent article \cite{Greensite:prep}, the survival of electric flux tubes at $T>T_c$, in the context of the gluon chain model \cite{Greensite:2001nx}, may also be connected with the persistence (indeed, the {\it increase}) of the color Coulomb string tension in the deconfined phase.

    We have also seen that the mere fact of positivity, of adjoint torelon loops at low temperatures, implies that the eigenvalues of fundamental representation Polyakov loops tend to slightly attract, rather than repel, in the confinement phase.  If this small deviation from the Haar measure reflects the nature of the underlying confinement mechanism, then confining configurations associated with center elements, i.e.\ center vortices, would seem to be favored.

\acknowledgments{J.G.'s research is supported in part by
the U.S.\ Department of Energy under Grant No.\ DE-FG03-92ER40711.  \v{S}.O. is supported in part by the Slovak Grant Agency for Science, Project VEGA No.\ 2/0070/09, by ERDF OP R\&D, Project CE meta-QUTE ITMS 26240120022, and via CE SAS QUTE. }

\bibliography{torlon}

\end{document}